\documentclass[a4paper,USenglish,cleveref]{lipics-v2021}
\author{Claire Mathieu}{CNRS, IRIF, Universit\'e de Paris, France \and \url{https://www.irif.fr/~claire/}}{claire.mathieu@irif.fr}{}{}
\author{Hang Zhou}{École Polytechnique, Institut Polytechnique de Paris, France \and \url{http://www.normalesup.org/~zhou/}}{hzhou@lix.polytechnique.fr}{}{}
\authorrunning{C. Mathieu and H. Zhou}
\Copyright{Claire Mathieu and Hang Zhou}

\funding{This work was partially funded by the grant ANR-19-CE48-0016 from the French National Research Agency (ANR).}
\relatedversion{An extended abstract of this paper appears in the proceedings of the International Symposium on Algorithms and Computation (ISAAC) 2021.}
\hideLIPIcs


\EventEditors{Hee-Kap Ahn and Kunihiko Sadakane}
\EventNoEds{2}
\EventLongTitle{32nd International Symposium on Algorithms and Computation (ISAAC 2021)}
\EventShortTitle{ISAAC 2021}
\EventAcronym{ISAAC}
\EventYear{2021}
\EventDate{December 6--8, 2021}
\EventLocation{Fukuoka, Japan}
\EventLogo{}
\SeriesVolume{212}
\ArticleNo{62}

\nolinenumbers 

\newtheorem{fact}[theorem]{Fact}
\newcommand{\E}{\mathbb{E}}
\newcommand{\eps}{\epsilon}
\renewcommand{\P}{\mathbb{P}}

\DeclareMathOperator{\erf}{erf}
\DeclareMathOperator{\cost}{cost}

\newcommand{\rad}{\mathrm{rad}}
\newcommand{\TSP}{\mathrm{TSP}}
\newcommand{\ITP}{\mathrm{ITP}}
\newcommand{\SP}{\hat w}
\newcommand{\mix}{\mathrm{mix}}

\newcommand{\I}{\mathrm{I}}
\newcommand{\II}{\mathrm{II}}

\newcommand{\OPT}{\mathrm{OPT}}

\title{Probabilistic Analysis of Euclidean Capacitated Vehicle Routing}

\begin{document}

\maketitle

\begin{abstract}
We give a probabilistic analysis of the unit-demand Euclidean capacitated vehicle routing problem in the random setting, where the input distribution consists of $n$ unit-demand customers modeled as independent, identically distributed uniform random points in the two-dimensional plane.
The objective is to visit every customer using a set of routes of minimum total length, such that each route visits at most $k$ customers, where $k$ is the capacity of a vehicle.
All of the following results are in the random setting and hold asymptotically almost surely.

The best known polynomial-time approximation for this problem is the iterated tour partitioning (ITP) algorithm, introduced in 1985 by Haimovich and Rinnooy Kan~\cite{haimovich1985bounds}.
They showed that the ITP algorithm is near-optimal when $k$ is either $o(\sqrt{n})$ or $\omega(\sqrt{n})$, and they asked whether the ITP algorithm was ``also effective in the intermediate range''.
In this work, we show that when $k=\sqrt{n}$, the ITP algorithm is at best a $(1+c_0)$-approximation for some positive constant $c_0$.

On the other hand, the approximation ratio of the ITP algorithm was known to be at most $0.995+\alpha$ due to Bompadre, Dror, and Orlin~\cite{bompadre2007probabilistic}, where $\alpha$ is the approximation ratio of an algorithm for the traveling salesman problem.
In this work, we improve the upper bound on the approximation ratio of the ITP algorithm to $0.915+\alpha$.
Our analysis is based on a new lower bound on the optimal cost for the metric capacitated vehicle routing problem, which may be of independent interest.
\end{abstract}

\section{Introduction}
\subparagraph*{Unit-Demand Euclidean CVRP.} In the \emph{capacitated vehicle routing problem (CVRP)}, we are given a set of $n$ \emph{customers} and a \emph{depot}.
There is an unlimited number of identical vehicles, each of an integer \emph{capacity} $k$.
The \emph{route} of a vehicle starts at the depot and returns there after visiting at most $k$ customers.
The objective is to visit every customer, using a set of routes of minimum total length.
Vehicle routing is a basic type of problems in operations research, and several books~(see~\cite{anbuudayasankar2016models,crainic2012fleet,golden2008vehicle,toth2002vehicle} among others) have been written on those problems.
We study the \emph{unit-demand Euclidean} version of the problem, in which each customer has unit demand, all locations (the customers and the depot) lie in the two-dimensional plane, and distances are given by the Euclidean metric.
The unit-demand Euclidean CVRP is a generalization of the Euclidean traveling salesman problem and is known to be NP-hard for all $k\geq 3$ (see~\cite{asano1997covering}).
Unless explicitly mentioned, all CVRP instances in this paper are assumed to be unit-demand Euclidean.

\subparagraph*{ITP Algorithm.} The best known polynomial-time approximation for the CVRP is a very simple algorithm, called \emph{iterated tour partitioning (ITP)}. This algorithm first computes a traveling salesman tour (ignoring the capacity constraint) using some other algorithm, then partitions the tour into segments such that the number of customers in each segment is at most $k$, and finally connects the endpoints of each segment to the depot so as to make a tour.
The ITP algorithm was introduced and refined by Haimovich and Rinnooy~Kan~\cite{haimovich1985bounds} and Altinkemer and Gavish~\cite{altinkemer1990heuristics} in the 1980s.
Its performance is parameterized by the choice of traveling salesman tour :
the approximation ratio of the ITP algorithm is $1+\alpha$, where $\alpha$ is the approximation ratio of the algorithm used to compute the traveling salesman tour in the first step.
Since the Euclidean traveling salesman problem admits a \emph{polynomial-time approximation scheme (PTAS)} by Arora~\cite{arora1998polynomial} and Mitchell~\cite{mitchell1999guillotine}, $\alpha$~can be set to any constant strictly greater than 1.

\subparagraph*{Random Setting.} Given the difficult challenges posed by the CVRP, researchers turned to an analysis beyond worst case, by making some probabilistic assumptions on the distribution of the input instance. In 1985, Haimovich and Rinnooy~Kan~\cite{haimovich1985bounds} gave the first probabilistic analysis on the ITP algorithm for the CVRP, where the customers are \emph{independent, identically distributed (i.i.d.)} random points.
An event $\mathcal{E}$ occurs \emph{asymptotically almost surely (a.a.s.)} if $\lim_{n\to\infty}\P[\mathcal{E}]=1$.
They showed that, the ITP algorithm is a.a.s.\ an $(\alpha+o(1))$-approximation for the CVRP when $k$ is either $o(\sqrt{n})$ or $\omega(\sqrt{n})$.\footnote{As observed in~\cite{haimovich1985bounds}, a solution of the ITP algorithm consists of two types of costs: the \emph{radial cost} and the \emph{local cost}.
When $k$ is $o(\sqrt{n})$ or $\omega(\sqrt{n})$, one of the two types dominates, the reason for which the solution is an $(\alpha+o(1))$-approximation (or even a $(1+o(1))$-approximation for the case of $k=o(\sqrt{n})$).}
The performance of the ITP algorithm in the intermediate range of $k=\Theta(\sqrt{n})$ was unknown.
They asked in~\cite{haimovich1985bounds} whether the ITP algorithm was ``also effective in the intermediate range''.

In our work, we study this question raised by Haimovich and Rinnooy~Kan~\cite{haimovich1985bounds}.
We give a probabilistic analysis of the ITP algorithm when the points are i.i.d.\ random, with a focus on the range of $k=\Theta(\sqrt{n})$.
Our first main result is a lower bound: even in the random setting, the ITP algorithm is at best a $(1+c_0)$-approximation a.a.s., for some constant $c_0>0$ (\cref{thm:not-ptas}), see \cref{sec:not-ptas}.

\begin{theorem}
\label{thm:not-ptas}
Consider the iterated tour partitioning algorithm for the unit-demand Euclidean capacitated vehicle routing problem.
Let $V$ be a set of $n$ i.i.d.\ uniform random points in $[0,1]^2$.
Let $k=\sqrt{n}$.
For some fixed depot $O\in \mathbb{R}^2$, there exists a constant $c_0>0$, such that, for any constant $\alpha>1$, there exists an $\alpha$-approximate traveling salesman tour on $V\cup\{O\}$, such that the approximation ratio of the algorithm is at least $1+c_0$ asymptotically almost surely.
\end{theorem}

\begin{remark*}
The $\alpha$-approximate traveling salesman tour in \cref{thm:not-ptas} is constructed using Karp's partitioning algorithm~\cite{karp1977probabilistic}.
\end{remark*}

On the other hand, the approximation ratio of the ITP algorithm is at most $1+\alpha$ due to Altinkemer and Gavish~\cite{altinkemer1990heuristics}.
In 2007, this ratio was improved by Bompadre, Dror, and Orlin~\cite{bompadre2007probabilistic} to  $0.995+\alpha$, a.a.s., when the points are i.i.d.\ uniform random in the unit square.\footnote{The analysis in~\cite{bompadre2007probabilistic} focused on the case of $\alpha=1$, though that analysis can be easily generalized to any $\alpha\geq 1$.
Bompadre, Dror, and Orlin~\cite{bompadre2007probabilistic} noted in their work that a ratio of $0.985+\alpha$ is achievable without giving the proof.
}
Here, using a different approach (Theorem~\ref{lem:ITP-upper-bound}),
we further improve the upper bound on the approximation ratio in this random setting to $0.915+\alpha$, a.a.s.\ (\cref{thm:better-ratio}), see \cref{sec:better-ratio}.
We generalize our results to multiple depots in \cref{sec:multi-depot}.

\begin{theorem}
\label{thm:better-ratio}
Consider the iterated tour partitioning algorithm for the unit-demand Euclidean capacitated vehicle routing problem.
Let $V$ be a set of $n$ i.i.d.\ uniform random points in $[0,1]^2$.
Let $k$ be any integer in $[1,n]$.
Let the depot $O$ be any point in $\mathbb{R}^2$.
For any constant $\alpha\geq 1$ and any $\alpha$-approximate traveling salesman tour on $V\cup\{O\}$, the approximation ratio of the algorithm is at most $0.915+\alpha$ asymptotically almost surely.
\end{theorem}

\subsection{Other Related Work}
\label{sec:related-work}
\subparagraph*{PTAS and Quasi-PTAS Results for the CVRP.}
Despite the difficulty of the CVRP, there has been progress on several special cases. A series of papers designed PTAS algorithms for small $k$:
 work by Haimovich and Rinnooy Kan~\cite{haimovich1985bounds}, when $k$ is constant; by
 Asano et al.~\cite{asano1997covering}  extending techniques in~\cite{haimovich1985bounds}, for $k=O(\log n/\log\log n)$; and by
Adamaszek, Czumaj, and Lingas~\cite{adamaszek2010ptas}, when  $k\leq 2^{\log^{f(\eps)}(n)}$.
For higher dimensional Euclidean metrics, Khachay and Dubinin~\cite{khachay2016ptas} gave a PTAS for fixed dimension $\ell$ and $k=O(\log^{\frac{1}{\ell}}(n))$.
For unbounded $k$, Das and Mathieu~\cite{das2015quasipolynomial} designed a quasi-polynomial time approximation scheme.

\subparagraph*{Probabilistic Analyses.} The instance distribution  when the customers are i.i.d. random points is perhaps the most natural probabilistic setting. In that setting,  Rhee~\cite{rhee1994probabilistic} and Daganzo~\cite{daganzo1984distance} analyzed the value of an optimal solution to the CVRP for the case when $k$ is fixed.
Baltz et al~\cite{baltz2007probabilistic} studied the multiple depot vehicle routing problem when both the customers and the depots are i.i.d.\ random points and assuming unlimited tour capacity.

\subparagraph*{Analyses of the ITP Algorithm.} Because of the popularity of the ITP algorithm, its approximation ratio has already been much studied and bounds were utilized in a design of best-to-date approximation algorithms for the CVRP, see, e.g.,~\cite{bompadre2006improved}.
In the metric version of the CVRP, the approximation ratio of the ITP algorithm is at most $1+(1-\frac{1}{k})\alpha$ due to Altinkemer and Gavish~\cite{altinkemer1990heuristics}.
Bompadre, Dror, and Orlin~\cite{bompadre2006improved} reduced this bound by a factor of $\Omega(\frac{1}{k^3})$.
On the other hand, Li and Simchi-Levi~\cite{li1990worst} showed that the ITP algorithm is at best a $(2-\frac{1}{k})$-approximation algorithm on general metrics even if $\alpha=1$.
Despite of a huge amount of research, the ITP algorithm by  Haimovich and Rinnooy Kan~\cite{haimovich1985bounds} and Altinkemer and Gavish~\cite{altinkemer1990heuristics} remains the polynomial-time algorithm with the best approximation guarantee for the Euclidean CVRP.

\subparagraph*{Other Applications of the ITP Algorithm.}
Very recently, Blauth, Traub, and Vygen~\cite{blauth2021improving} exploited properties of tight instances in the analysis of the ITP algorithm, and used those properties in their design of the best-to-date approximation algorithm for metric CVRP with a ratio of $1+\alpha-\eps$, where $\eps$ is roughly $\frac{1}{3000}$.

Because of its simplicity, the ITP algorithm is versatile and has been adapted to other vehicle routing problems.
For example, Mosheiov~\cite{mosheiov1998vehicle} studied the vehicle routing with pick-up and delivery services.
They showed that the ITP algorithm is efficient through worst-case analysis and numerical tests.
Li, Simchi-Levi, and Desrochers~\cite{li1992distance} considered the vehicle routing problem with constraints on the total distance traveled by each vehicle.
They showed that the ITP algorithm has a good worst-case performance when the number of vehicles is relatively small.

\subsection{Overview of Techniques}

To show that the ITP algorithm is at best a $(1+c_0)$-approximation for the CVRP in the random setting (\cref{thm:not-ptas}), we construct a significantly better solution.

In the random setting, one may view an ITP solution as partitioning the unit square into small regions and dedicating one tour to each small region. The cost of the solution is then roughly the sum of two terms: the \emph{radial cost}, incurred by traveling between the depot and the small region; and the \emph{local cost}, incurred by traveling  from customer to customer within the small region.

To improve that solution, the idea is that instead of traveling straight between the depot and the small region, a smarter tour  might as well make some small detours to visit some additional nearby customers en route to the small region. We call that a \emph{mixed tour}. This modification of the solution has a positive effect because those nearby customers are covered at little additional cost, thus saving the local cost of covering those customers; but it also has a negative effect because visiting those nearby customers uses up some of the tour's capacity, and to account for that the definition of the small regions must be adjusted, and their area shrunk. Controlling the two competing effects so that on balance the net result is an improvement requires a delicate definition of regions.
We start by decomposing the plane into regions of three types.
Then we construct a solution in which a single tour may visit regions of different types, see~\cref{fig:decomposition}.
The mixed structure of the tours enables us to show that the constructed solution has significantly smaller cost. See \cref{sec:not-ptas} for more details.

Our proof of the improved upper bound on the approximation ratio of the ITP algorithm in the random setting (Theorem~\ref{thm:better-ratio}) relies on a new lower bound on the optimal cost (\cref{thm:new-lower-bound}).
To achieve the new lower bound, we consider the gap between the average distance to the depot and the maximum distance to the depot among all points in a single tour of an optimal solution.
Intuitively, if this gap is large, then the gap itself contributes to the lower bound on the optimal cost; and if this gap is small, then there are many points whose distances to the depot is close to the maximum distance, and the total local cost of those points contributes to the lower bound.
Our analysis for the lower bound is completely different from~\cite{bompadre2007probabilistic}  and enables us to obtain a better approximation ratio of $0.915+\alpha$.

Our new lower bound on the optimal cost also enables us to generalize our results to the setting of multiple depots.
This lower bound holds in the metric CVRP in general, and may be of independent interest.

\begin{remark*}
The restriction to i.i.d.\ uniform random points in $[0,1]^2$ is made to simplify the presentation.
With extra work, our analysis can be extended to higher dimensional Euclidean spaces, to general density functions, and to general bounded supports (though the approximation guarantees in those settings may differ from that in \cref{thm:better-ratio}).
\end{remark*}

\section{Notations and Preliminaries}

Let $\delta(\cdot,\cdot)$ denote the Euclidean distance between two points or between a point and a set of points. 
For any path $P$ of points $x_1,x_2,\dots, x_m$ in $\mathbb{R}^2$ where $m\in \mathbb{N}$, define $\cost(P)=\sum_{i=1}^{m-1} \delta(x_i,x_{i+1})$.

\subparagraph*{Capacitated Vehicle Routing Problem (CVRP).}
Given a set $V$ of $n$ points in $\mathbb{R}^2$, a depot $O$ in $\mathbb{R}^2$, and an integer capacity $k\in[1,n]$, the goal is to find a collection of tours covering $V$ of minimum total cost, such that each tour visits $O$ and at most $k$ points in $V$.
Let $\OPT$ denote the value of an optimal solution to the CVRP.

For any point $x\in V$, let $\ell(x)=\delta(O,x)$.
Let $\rad$ denote the \emph{radial cost}, defined by $\rad=\frac{2}{k}\cdot \sum_{x\in V} \ell(x).$

\begin{lemma}[\cite{haimovich1985bounds}]
\label{lem:OPT-lower-bound}
Let $T^*$ be an optimal traveling salesman tour on $V\cup \{O\}$.
Then $\OPT\geq \max(\rad,\cost(T^*)).$
\end{lemma}

\subparagraph*{Iterated Tour Partitioning (ITP).}
We review the \emph{iterated tour partitioning (ITP)} algorithm defined by Altinkemer and Gavish~\cite{altinkemer1990heuristics}.
The ITP algorithm consists of a preprocessing phase and a main phase.
In the preprocessing phase, it runs an approximation algorithm for the traveling salesman problem   on $V\cup \{O\}$.
Let $\alpha$ denote the approximation ratio of this algorithm.
Let $T=(O,x_1,x_2,\dots,x_n,O)$ denote the resulting traveling salesman tour.
In the main phase, the ITP algorithm selects the best of the $k$ solutions constructed as follows.
For each $i\in[1,k]$, let $n_i=\lceil (n-i)/k\rceil +1$ and define a solution $S_i$ to the CVRP to be the union of the $n_i$ tours $(O,x_1,\dots,x_i,O)$, $(O, x_{i+1},\dots, x_{i+k},O)$, $(O, x_{i+k+1}, \dots, x_{i+2k},O)$, \dots, $(O, x_{i+(n_i-2)k+1}, \dots, x_{n}, O)$.
In other words, the solution $S_i$ partitions the traveling salesman tour $T$ into segments with $k$ points each, except possibly the first and the last segments.
The output of the ITP algorithm is a solution among $S_1,\dots,S_k$ that achieves the minimum cost.
It is easy to see that the main phase of the ITP algorithm can be carried out in $O(nk)$ time.\footnote{The running time of the main phase can even be improved to $O(n)$.}

Let $\ITP(T)$ denote the cost of the output solution.
The following classic bound on $\ITP(T)$ was due to Altinkemer and Gavish~\cite{altinkemer1990heuristics} and, together with Lemma~\ref{lem:OPT-lower-bound}, immediately implies that the ITP algorithm is a $(1+\alpha)$-approximation, where $\alpha$ is the approximation ratio of the traveling salesman tour $T$.

\begin{lemma}[\cite{altinkemer1990heuristics}]
\label{lem:ITP-upper-bound}
Let $T$ be any traveling salesman tour on $V\cup \{O\}$.
Then \[\OPT\leq \ITP(T)\leq \rad+\left(1-\frac{1}{k}\right)\cdot\cost(T).\]
\end{lemma}

\subparagraph*{Probabilistic Analysis of the Traveling Salesman Problem.}
Beardwood, Halton, and Hammersley~\cite{beardwood1959shortest} analyzed the value of an optimal solution to the traveling salesman problem in the random setting.
\begin{lemma}[\cite{beardwood1959shortest,steinerberger2015new}]
\label{lem:TSP}
Let $V$ be a set of $n$ i.i.d.\ uniform random points with bounded support in~$\mathbb{R}^2$.
Let $M$ denote the measure of the support.
Let $T^*$ denote an optimal traveling salesman tour on $V$.
Then there exists a universal constant $\beta$ such that, for any $\eps>0$, we have \[\lim_{n\to\infty}\frac{\cost(T^*)}{\sqrt{M\cdot n}}=\beta,\quad\text{with probability 1.}\]
In addition, $\beta_0< \beta<\beta_1$, where $\beta_0=0.62866$ and $\beta_1=0.92117$.
\end{lemma}

\begin{remark*}
Up to scaling, \cref{lem:TSP} holds for any support that is a rectangle with constant aspect ratio.
\end{remark*}

\section{Lower Bound on the Approximation Ratio}
\label{sec:not-ptas}

In this section, we prove \cref{thm:not-ptas} by providing a lower bound on the approximation ratio  $\ITP(T)/\OPT$ of the ITP algorithm, where $T$ is a traveling salesman tour.
Let the depot $O=\left(\frac{1}{2}, -1000\right)$.

\cref{lem:OPT-better,lem:ITP-tight} are central to the proof of \cref{thm:not-ptas} and contain main novelties in this section.

\begin{lemma}
\label{lem:OPT-better}
Let $\beta$ be defined as in Lemma~\ref{lem:TSP}. Then there exists a constant $c_1\in(0,\beta)$ such that for any $\eps_1>0$, $\OPT<(1+\eps_1)(\rad+\beta \sqrt{n})-c_1\sqrt{n}$, a.a.s.
\end{lemma}

\begin{lemma}
\label{lem:ITP-tight}
Let $\beta$ be defined as in Lemma~\ref{lem:TSP}. Then for any $\alpha>1$, there exists an $\alpha$-approximate traveling salesman tour $T$ on $V\cup\{O\}$, such that for any $\eps_1>0$, $\ITP(T)>(1-\eps_1)(\rad+\beta \sqrt{n})$, a.a.s.
\end{lemma}
In the rest of the section, we prove \cref{lem:OPT-better,lem:ITP-tight} in \cref{sec:proof-OPT-better,sec:proof-ITP-tight}, respectively.
Finally, we give the proof of \cref{thm:not-ptas} in \cref{sec:proof-not-ptas}.

\subsection{Proof of \cref{lem:OPT-better}}
\label{sec:proof-OPT-better}
Without loss of generality, we assume that $\eps_1\leq 1$, since otherwise it suffices to prove the claim for the case of $\eps_1=1$.
We construct a solution to the CVRP whose cost is less than $(1+\eps_1)(\rad+\beta \sqrt{n})-c_1\sqrt{n}$, a.a.s., where the constant $c_1>0$ will be chosen later.

\subsubsection{Decomposition of the Plane}
\begin{figure}[t]
     \centering
     \begin{subfigure}[b]{0.3\textwidth}
         \centering
         \includegraphics[scale=0.7]{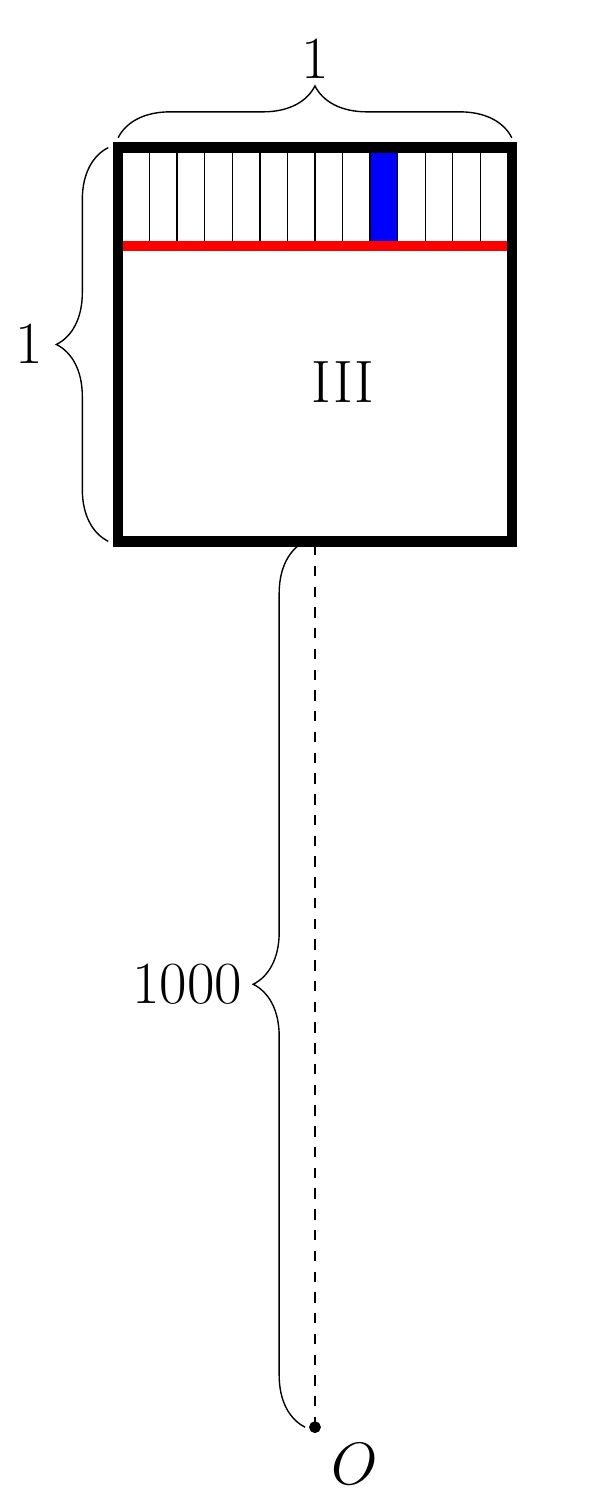}
         \caption{$[0,1]^2$ decomposition.}
        \label{fig:square}
     \end{subfigure}
     \hfill
     \begin{subfigure}[b]{0.35\textwidth}
         \centering
         \includegraphics[scale=0.7]{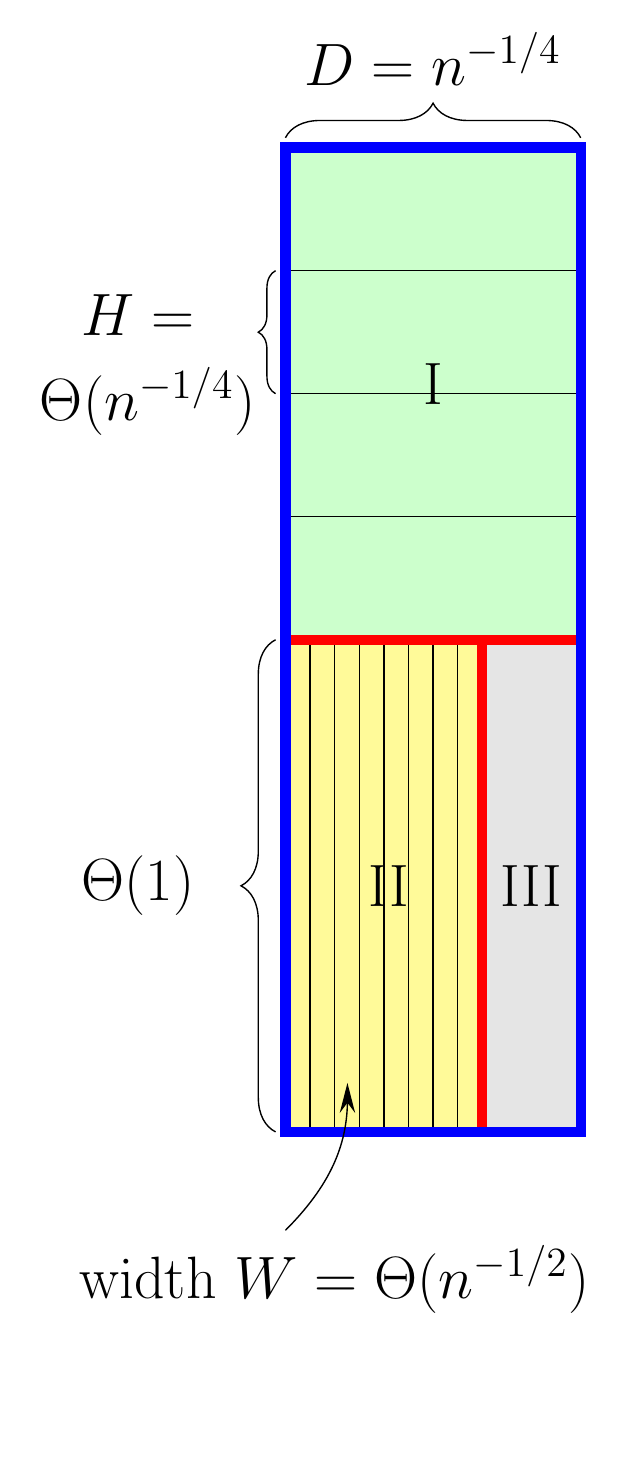}
         \caption{Box decomposition.}
         \label{fig:box}
     \end{subfigure}
     \hfill
     \begin{subfigure}[b]{0.33\textwidth}
         \centering
         \includegraphics[scale=0.7]{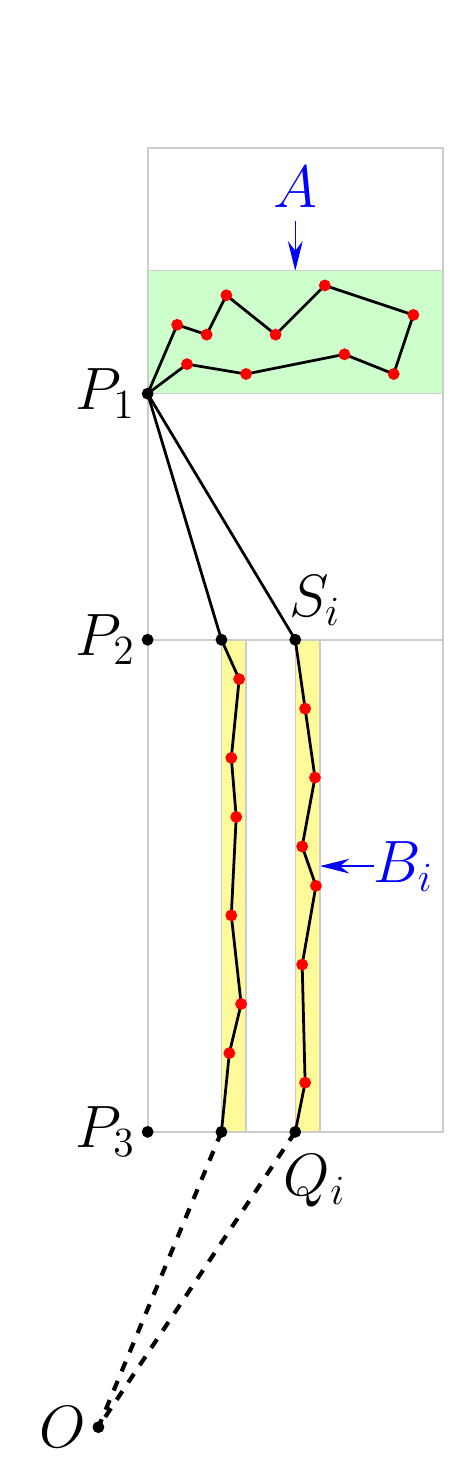}
         \caption{A mixed tour.}
         \label{fig:tour}
     \end{subfigure}
        \caption{Decomposition and tour construction. \cref{fig:square} illustrates  the decomposition of $[0,1]^2$. The highlighted area in \cref{fig:square} represents a box.
        \cref{fig:box} illustrates the decomposition of a box into rectangles of  types I, II, and III.
        \cref{fig:tour} describes a tour covering points in one rectangle $A$ of type I and in two rectangles $B_i$, for $i\in\{1,2\}$, of type II. }
       \label{fig:decomposition}
\end{figure}

In order to construct a solution to the CVRP, we describe a decomposition of $[0,1]^2$ into rectangles of three types.
Let $\eps_2=\frac{\eps_1}{10}$.
We partition\footnote{The decomposition is a partition except for the boundaries, that have measure 0.} $[0,1]^2$ into a lower part  $[0,1]\times[0,\frac{3+\eps_2}{4}]$  which is a rectangle of type III, and a collection of \emph{boxes} of the form $ [(i-1)D,iD]\times [\frac{3+\eps_2}{4},1]$, with $D=n^{-1/4}$ and $1\leq i\leq 1/D$.
For simplicity, assume that $1/D$ is an integer.
See \cref{fig:square}.

Next, we decompose each box, see \cref{fig:box}. Let $m=\frac{5}{40-\beta}\cdot n^{1/4}$. For simplicity, assume that $m$ is an integer. The upper half of a box is partitioned into  $m$ \emph{type I rectangles}  of the form $[(i-1)D,iD]\times [1-(j-1)H, 1-jH]$, where $H=\frac{1-\eps_2}{8\cdot m}$ and $1\leq j\leq m$. The lower left part of a box is partitioned into $2m$ slices such that each slice is a \emph{type II rectangles} of the form $[(i-1)D+(j-1)W, (i-1)D+jW]\times [\frac{3+\eps_2}{4},\frac{7+\eps_2}{8}]$, with $W=\frac{\beta}{10}\cdot n^{-1/2}$ and $1\leq j\leq 2m$. The rest of the box is a single \emph{type III rectangle}.

For any rectangle $R$ in the resulting decomposition, let $n_R$ denote the number of points of $V$ that are in the rectangle $R$, and let $M_R$ denote the measure of the rectangle~$R$.
The following fact relates $n_R$ with $M_R$.
\begin{fact}
\label{fact:eventE}
A.a.s.  the following event $\mathcal{E}$ occurs: $(1-\eps_2)\cdot M_R \cdot n <n_R<(1+\eps_2)\cdot M_R \cdot n$ for all rectangles $R$ in the resulting decomposition.
\end{fact}

\begin{proof}
Let $R$ be any rectangle in the resulting decomposition.
Observe that $M_R=\Omega(1/\sqrt{n})$.
The expectation of $n_R$ is $M_R\cdot n=\Omega(\sqrt{n})$.
By Chernoff bound,\[\P\Big[n_R\leq (1-\eps_2)M_R\cdot n\Big]\leq e^{-\Omega(\sqrt{n})} \quad\text{and}\quad\P\Big[n_R\geq (1+\eps_2)M_R\cdot n\Big]\leq e^{-\Omega(\sqrt{n})}.\]
Since there are $\Theta(\sqrt{n})$ rectangles in the decomposition, the event $\mathcal{E}$ occurs with probability at least $1-\Theta(\sqrt{n}) \cdot e^{-\Omega(\sqrt{n})}=1-o(1)$.
\end{proof}

From now on, we condition on the occurrence of $\mathcal{E}$ in \cref{fact:eventE}.

\subsubsection{Construction of a Solution}
\label{sec:construction}
To construct a relative cheap solution, the main observation is that it is profitable for a tour to visit points in rectangles of both types I and II.
In each box, there are $m$ rectangles of type I and $2m$ rectangles of type II. We form $m$ groups with those, such that each group contains one rectangle of type I and two rectangles of type II. For each group, we cover the points in the group by  a particular tour on these points in addition to the depot $O$, in a way to be described shortly.
For all points in the rectangles of type III, we construct an optimal solution to the CVRP on those points with depot $O$ and with capacity $\sqrt{n}$.

A \emph{mixed tour} is a tour covering points in rectangles of types I and II.
Consider a box $\mathcal B$ and a group in $\mathcal B$ consisting of a rectangle $A$ of type I and two rectangles $B_1$ and $B_2$ of type II.
We construct a specific mixed tour $T_{\mix}$ defined as follows.

Let $P_1$ denote the bottom left corner of the rectangle $A$.
Let $T_0$ denote an optimal traveling salesman tour on the points in $A\cup\{P_1\}$.
For each $i\in\{1,2\}$, we define an $O$-to-$P_1$ path $T_i$ visiting the points in $B_i$ as follows.
Let $S_i$ and $Q_i$ denote the top left and bottom left corners of the rectangle $B_i$.
Let $T_i$ denote the concatenation of the segment $OQ_i$, a $Q_i$-to-$S_i$ path visiting the points in $B_i$ in non-decreasing order on the $y$-coordinate (breaking ties arbitrarily), and the segment $S_iP_1$.
Finally, the tour $T_{\mix}$ is defined as the concatenation of $T_0$, $T_1$, and $T_2$.
See \cref{fig:tour}. This completes our construction.

Since the measure of $A$ is $D\cdot H$ and the measure of $B_i$ (for each $i\in\{1,2\}$) is $\frac{1-\eps_2}{8}\cdot W$, Event $\mathcal{E}$ implies that the total number of points in $A\cup B_1\cup B_2$ is at most \[(1+\eps_2)\left(D\cdot H+2\cdot \frac{1-\eps_2}{8}\cdot W\right)\cdot n=(1+\eps_2)(1-\eps_2)n^{-1/2}\cdot n<\sqrt{n},\]
so the constructed solution is feasible.

\subsubsection{Cost of a Mixed Tour}
Consider any mixed tour $T_{\mix}$.
We follow the same notations as in \cref{sec:construction}.
From the construction,
\begin{equation}
\label{eqn:T-mix-def}
\cost(T_{\mix})=\cost(T_0)+\cost(T_1)+\cost(T_2).
\end{equation}

Let $T_A^*$ denote an optimal traveling salesman tour on the points in $A$.
The cost of $T_0$ is at most $\cost(T_A^*)$ plus the cost of the detour to include the point $P_1$.
The cost of the detour is less than $2(D+H)$, so
\begin{equation}
\label{eqn:T0}
\cost(T_0)<\cost(T_A^*)+2(D+H).
\end{equation}

Let $n_1$ and $n_2$ denote the number of points of $V$ that are in $B_1$ and $B_2$, respectively.
The costs of $T_1$ and $T_2$ are bounded by the following fact.
\begin{fact}
\label{fact:Ti}
For $i\in\{1,2\}$, we have
\begin{equation}
\label{eqn:Ti}
\cost(T_i)< \delta(O,P_1)+\frac{1}{4000}+n_i W+(W+2D).
\end{equation}
\end{fact}

\begin{proof}
Let $T_i'$ denote the subpath of $T_i$ that is between $Q_i$ and $S_i$.
We have \[\cost(T_i)= \delta(O,Q_i)+\cost(T_i')+\delta(S_i,P_1).\]
To analyze $\cost(T_i')$, recall that the Euclidean length of a segment is at most its length in the $\ell_1$ metric. Since $T_i'$ is monotone in the $y$-coordinate and stays within a rectangle of width $W$, its cost is at most $\delta(Q_i,S_i)+(n_i+1) W$.
We have
\begin{equation*}
\cost(T_i)\leq \delta(O,Q_i)+\delta(Q_i,S_i)+(n_i +1)W+\delta(S_i,P_1).
\end{equation*}

Let $P_2$ denote the midpoint of the left boundary of the box $\mathcal B$, and let $P_3$ denote the bottom left corner of the box $\mathcal{B}$.
We have $\delta(Q_i,S_i)=\delta(P_3,P_2)$ by the construction.
Since the box $\mathcal{B}$ has width $D$, we have $\delta(O,Q_i)\leq \delta(O,P_3)+D$ and $\delta(S_i,P_1)\leq \delta(P_2,P_1)+D$ by the triangle inequality.
Using $\delta(P_3,P_2)+\delta(P_2,P_1)=\delta(P_3,P_1)$, we have
\begin{equation}
\label{eqn:cost-Ti-P3}
\cost(T_i)\leq \delta(O,P_3)+\delta(P_3,P_1)+(n_i+1)W+2D.
\end{equation}

To analyze $\delta(O,P_3)+\delta(P_3,P_1)$, let $P_3'$ denote the point on the segment $OP_1$ that has the same $y$ coordinate as $P_3$.
By the triangle inequality, $\delta(O,P_3)\leq \delta(O,P_3')+\delta(P_3,P_3')$ and $\delta(P_3,P_1)\leq \delta(P_3',P_1)+\delta(P_3,P_3')$.
Using $\delta(O,P_3')+\delta(P_3',P_1)=\delta(O,P_1)$, we have
\begin{equation}
\label{eqn:O-P3}
\delta(O,P_3)+\delta(P_3,P_1)\leq \delta(O,P_1)+2\delta(P_3,P_3').
\end{equation}

It remains to bound $\delta(P_3,P_3')$.
Let $\Delta_x$ (resp.\ $\Delta_y$) denote the absolute difference of the $x$-coordinate (resp.\ $y$-coordinate) between $O$ and $P_3$.
We observe that
\[\delta(P_3,P_3')=\Delta_x\cdot\frac{\delta(P_3,P_1)}{\Delta_y+\delta(P_3,P_1)}.\]
By the definition of a box, $\delta(P_3,P_1)<\frac{1}{4}$.
Since $O=\left(\frac{1}{2},-1000\right)$ and $P_3\in[0,1]^2$, we have $\Delta_x\leq \frac{1}{2}$ and $\Delta_y\geq 1000$.
Thus
\begin{equation}
\label{eqn:P3-P3'}
\delta(P_3,P_3')<\frac{1}{8000}.
\end{equation}
From \cref{eqn:cost-Ti-P3,eqn:O-P3,eqn:P3-P3'}, we have
\[\cost(T_i)< \delta(O,P_1)+\frac{1}{4000}+n_i W + (W+2D).\]
The claim follows.
\end{proof}

From \cref{eqn:T-mix-def,eqn:T0,eqn:Ti}, and using the definition of $W$, we have
\begin{equation}
\label{eqn:T-mix-1}
\cost(T_{\mix})<\cost(T_A^*)+2\delta(O,P_1)+\frac{1}{2000}+\frac{\beta}{10}\cdot\frac{n_1+n_2}{\sqrt{n}}+(2W+6D+2H).
\end{equation}
It remains to bound $\delta(O,P_1)$.
Observe that by the definition of $\ell(\cdot)$ and the triangle inequality, and since the height of a box $\mathcal{B}$ is less than $\frac{1}{4}$,
\[\delta(O,P_1)<
\left\{ \begin{array}{ll}
\ell(x)+D+H& \quad\text{for any }x\in A,\\
\ell(x)+\frac{1}{4}+D+H &\quad\text{for any }x\in B_1\cup B_2.
\end{array}
\right. \]
Let $V_{\mix}$ denote the set of the points of $V$ in $A\cup B_1\cup B_2$.
By averaging we have
\begin{equation*}
\delta(O,P_1)<\frac{1}{|V_{\mix}|}\left(\sum_{x\in V_{\mix}}\ell(x)\right)+\frac{1}{|V_{\mix}|}\frac{n_1+n_2}{4}+ (D+H).
\end{equation*}
Since the measure of $A\cup B_1\cup B_2$ is $(1-\eps_2)n^{-1/2}$, Event $\mathcal{E}$ implies that $|V_{\mix}|>(1-\eps_2)^2\cdot\sqrt{n}$, which is at least $\frac{\sqrt{n}}{1+\eps_1}$ since $\eps_2=\frac{\eps_1}{10}$.
Hence
\begin{equation}
\label{eqn:P1}
\delta(O,P_1)<\frac{1+\eps_1}{\sqrt{n}}\left[\left(\sum_{x\in V_{\mix}}\ell(x)\right)+\frac{n_1+n_2}{4}\right]+(D+H).
\end{equation}
Since $n_1+n_2=\Theta(\sqrt{n})$, we have $(2W+6D+2H)+2(D+H)<\frac{\eps_1}{\sqrt{n}}\cdot \frac{\beta}{10}\cdot(n_1+n_2)$ when $n$ is large enough.
From \cref{eqn:T-mix-1,eqn:P1}, we have
\begin{equation}
\label{eqn:T-mix-2}
\cost(T_{\mix})<\cost(T_A^*)+\frac{1+\eps_1}{\sqrt{n}}\left[2\left(\sum_{x\in V_{\mix}}\ell(x)\right)+\left(\frac{\beta}{10}+\frac{1}{2}\right)(n_1+n_2)\right]+\frac{1}{2000}.
\end{equation}

\subsubsection{Cost of the Solution in Rectangles of Types I and II}
\label{sec:rect-I-II}
Let $V_{\I}$ and $V_{\II}$ denote the subsets of the points of $V$ that are in the rectangles of type I and type II, respectively.
Let $K$ denote the number of mixed tours.
We have $K=n^{1/4}\cdot m=\frac{5}{40-\beta}\cdot\sqrt{n}$.
Applying \cref{eqn:T-mix-2} on each mixed tour and summing, we have
\begin{equation}
\label{eqn:S-mix}
\cost(\mathcal{S}_{\mix})\leq Y+\frac{1+\eps_1}{\sqrt{n}}\left[2\left(\sum_{x\in V_{\I}\cup V_{\II}}\ell(x)\right)+\left(\frac{\beta}{10}+\frac{1}{2}\right)|V_{\II}|\right]+\frac{\sqrt{n}}{400\cdot\left(40-\beta\right)},\quad
\end{equation}
where $Y$ denotes the overall cost of $T_A^*$ over all rectangles $A$ of type I.

To analyze $Y$, we consider any rectangle $A$ of type I.
The measure of $A$ is $M_A=D\cdot H=(1-\eps_2)\left(1-\frac{\beta}{40}\right)\frac{1}{\sqrt{n}}$.
The event $\mathcal{E}$ implies that
\begin{equation}
\label{eqn:nA}
(1-\eps_2)^2 \cdot \left(1-\frac{\beta}{40}\right) \cdot \sqrt{n}<n_A<\left(1-\frac{\beta}{40}\right)\cdot \sqrt{n}.
\end{equation}
We investigate the expectation of $\cost(T_A^*)$.
By a construction given in \cite{beardwood1959shortest}, there exists a constant $C$ such that the cost of an optimal traveling salesman tour through any $n_A$ points in $A$ is at most $C\sqrt{M_A\cdot n_A}$.
Together with \cref{lem:TSP}, it follows that
\[\E[\cost(T_A^*)]<(1+\eps_2)\cdot\beta\sqrt{M_A\cdot n_A}<(1+\eps_2)\cdot \beta \left(1-\frac{\beta}{40}\right),\]
for $n$ large enough (and thus $n_A$ large enough).
Since there are $K=\Theta(\sqrt{n})$ rectangles $A$ of type I, by the law of large numbers,
\[Y=\sum_A \cost(T_A^*)< (1+\eps_2)\cdot K \cdot (1+\eps_2)\cdot \beta \left(1-\frac{\beta}{40}\right),\quad\text{a.a.s.}\]
On the other hand, summing \cref{eqn:nA} over all $A$, we have
\[|V_\I|=\sum_A n_A>K\cdot (1-\eps_2)^2 \cdot \left(1-\frac{\beta}{40}\right) \cdot \sqrt{n}.\]
Therefore,
\[Y<(1+\eps_2)^2\cdot \frac{1}{(1-\eps_2)^2}\cdot\frac{\beta\cdot |V_\I|}{\sqrt{n}}<(1+\eps_1)\cdot\frac{\beta\cdot |V_\I|}{\sqrt{n}},\quad\text{a.a.s.,}\]
where the last inequality follows since $\eps_2=\frac{\eps_1}{10}$.
From \cref{eqn:S-mix}, we have, a.a.s.,
\begin{equation}
\label{eqn:rect-I-II}
\cost(\mathcal{S}_{\mix})\leq \frac{1+\eps_1}{\sqrt{n}}\left[2\left(\sum_{x\in V_{\I}\cup V_{\II}}\ell(x)\right)+\beta\cdot|V_\I|+\left(\frac{\beta}{10}+\frac{1}{2}\right)|V_{\II}|\right]+\frac{\sqrt{n}}{400\cdot\left(40-\beta\right)}.
\end{equation}

\subsubsection{Cost of the Solution in Rectangles of Type III}
Let $\hat V$ denote the subset of the points of $V$ that are in the rectangles of types III.
Let $\hat T$ denote an optimal traveling salesman tour on $\hat V\cup\{O\}$.
Let $\hat{\mathcal{S}}$ denote an optimal solution to the CVRP on $\hat V$ with depot $O$ and with capacity $k=\sqrt{n}$.
By \cref{lem:ITP-upper-bound}, \[\cost(\hat{\mathcal{S}})\leq \frac{2}{\sqrt{n}}\bigg(\sum_{x \in \hat V}\ell(x)\bigg)+\cost(\hat T).\]
The cost of $\hat T$ is at most the cost $C_{\TSP}$ of an optimal traveling salesman tour on $\hat V$ plus the detour to visit $O$.
Since the distance between the depot and any point in $[0,1]^2$ is $O(1)$, the cost of the detour is $O(1)$, so $\cost(\hat T)\leq C_{\TSP}+O(1)$.

Next, we analyze $C_{\TSP}$.
Let $L$ denote the width of a type III rectangle inside a box.
Then $L=D-W\cdot m=\Theta(n^{-1/4})$.
We observe that the length of a side of any type III rectangle is either $L$ or $\omega(L)$.
So we partition every type III rectangle into squares of side length $L$.
For each square, consider an optimal traveling salesman tour on the points inside that square.
Let $Z$ denote the overall cost of the optimal traveling salesman tours inside all squares from all rectangles of type III.
Then $C_{\TSP}$ is at most $Z$ plus the total lengths of the boundaries of all squares.
Using the same argument from \cref{sec:rect-I-II}, we have
\[Z<\left(1+\frac{\eps_1}{3}\right)\cdot\frac{\beta\cdot|\hat V|}{\sqrt{n}},\quad\text{a.a.s.}\]
Since the boundary length of each square is negligible compared with the TSP cost inside that square, we have
\[C_{\TSP}=(1+o(1))\cdot Z <\left(1+\frac{\eps_1}{2}\right)\cdot\frac{\beta\cdot|\hat V|}{\sqrt{n}},\quad\text{a.a.s.}\]
Noting that $|\hat V|=\Theta(n)$, we have
\[\cost(\hat T)\leq C_{\TSP}+O(1)<(1+\eps_1)\cdot\frac{\beta\cdot|\hat V|}{\sqrt{n}},\quad\text{a.a.s.}\]
Therefore,
\begin{equation}
\label{eqn:rect-III}
\cost(\hat{\mathcal{S}})\leq \frac{2}{\sqrt{n}}\left(\sum_{x \in \hat V}\ell(x)\right)+\frac{1+\eps_1}{\sqrt{n}}\cdot\beta\cdot |\hat V|, \quad\text{a.a.s.}
\end{equation}

\subsubsection{Cost of the Global Solution}
Let $\mathcal{S}=\mathcal{S}_{\mix}\cup \hat{\mathcal{S}}$ denote the global solution.
From \cref{eqn:rect-III,eqn:rect-I-II}, and using $\frac{2}{\sqrt{n}}\cdot \sum_{x \in V}\ell(x)=\rad$ and $V_{\I}\cup V_{\II}\cup \hat V=V$, we have
\begin{equation}
\label{eqn:cost-S}
\cost(\mathcal{S})\leq (1+\eps_1)\left[\rad+\beta\cdot\frac{|V|}{\sqrt{n}}-\left(\beta-\frac{\beta}{10}-\frac{1}{2}\right)\frac{|V_{\II}|}{\sqrt{n}}\right]+\frac{\sqrt{n}}{400\cdot\left(40-\beta\right)},\quad\text{a.a.s.}
\end{equation}
Observe that the rectangles of type II have an overall measure of $(1-\eps_2)\cdot \frac{\beta}{8\cdot (40-\beta)}$.
Event $\mathcal{E}$ implies that $|V_{\II}|>(1-\eps_2)^2\cdot \frac{\beta \cdot n}{8\cdot (40-\beta)}>\frac{1}{1+\eps_1}\cdot \frac{\beta \cdot n}{8\cdot (40-\beta)}$ since $\eps_2=\frac{\eps_1}{10}$.
By \cref{lem:TSP}, $\beta>\beta_0=0.62866$.
Thus $\frac{9\beta}{10}-\frac{1}{2}>0$.
From \cref{eqn:cost-S}, we have, a.a.s,
\begin{align*}
(1+\eps_1)(\rad+\beta\sqrt{n})-\cost(\mathcal{S})
&\geq (1+\eps_1)\cdot\left(\frac{9\beta}{10}-\frac{1}{2}\right)\cdot \frac{|V_{\II}|}{\sqrt{n}}-\frac{\sqrt{n}}{400\cdot(40-\beta)}\\
&>\left(\frac{9\beta}{10}-\frac{1}{2}\right)\cdot \frac{\beta \sqrt{n}}{8\cdot (40-\beta)}-\frac{\sqrt{n}}{400\cdot(40-\beta)}\\
&=\frac{\sqrt{n}}{8\cdot (40-\beta)}\cdot \left(\left(\frac{9\beta}{10}-\frac{1}{2}\right)\cdot \beta-\frac{1}{50}\right)\\
&>\frac{\sqrt{n}}{8\cdot (40-\beta_0)}\cdot \left(\left(\frac{9\beta_0}{10}-\frac{1}{2}\right)\cdot \beta_0- \frac{1}{50}\right).
\end{align*}
Let $c_1$ denote the leading constant in the above bound, i.e.,
\[\displaystyle c_1=\frac{1}{8\cdot (40-\beta_0)}\cdot \left( \left(\frac{9\beta_0}{10}-\frac{1}{2}\right)\cdot \beta_0-\frac{1}{50}\right).\]
The value of $c_1$ is roughly 0.000068.
We conclude that
\[\cost(\mathcal{S})<(1+\eps_1)(\rad+\beta\sqrt{n})-c_1\sqrt{n}, \quad\text{a.a.s}.\]
We complete the proof of \cref{lem:OPT-better}.

\subsection{Proof of \cref{lem:ITP-tight}}
\label{sec:proof-ITP-tight}
Without loss of generality, we assume that $\alpha\leq 2$, since otherwise it suffices to prove the claim for the case of $\alpha=2$.
We also assume that $\eps_1\leq 1$ since otherwise the claim is evident.

To begin with, we construct a traveling salesman tour $T$ on $V\cup\{O\}$.
Let $\eps_0=\alpha-1$. Note that $\eps_0\in(0,1]$.
Let $\eps_2=\min\left(\frac{\eps_0}{20},\frac{\eps_1}{4}\right)$.
To avoid purely technical complications, we assume that $\frac{1}{\sqrt{1+\eps_2}}\cdot n^{1/4}$ is an integer.
We divide $[0,1]^2$ into squares of side length $D=\sqrt{1+\eps_2}\cdot n^{-1/4}$, resulting in $m=\frac{1}{D^2}=\frac{1}{1+\eps_2}\cdot \sqrt{n}$ squares.
Let $A_1,\dots,A_m$ be an ordering of the resulting squares, such that for each $i\in[1,m-1]$, the squares $A_i$ and $A_{i+1}$ are adjacent in the plane.
For each square $A_i$, let $V_i\subseteq V$ be the set of points of $V$ that are in the square $A_i$.
Let $T_i$ be an optimal traveling salesman tour on $V_i$.
Let $T$ be a traveling salesman tour on $V\cup\{O\}$ obtained by including $T_1, \dots, T_m$, adding a segment between the last point of $T_i$ and the first point of $T_{i+1}$ for each $i\in[1,m-1]$, and also adding a segment between the depot $O$ and the first point of $T_1$ (resp.\ the last point of $T_m$).

For any $i\in[1,m]$, the expectation of $|V_i|$ is $D^2 n=(1+\eps_2)\sqrt{n}$.
Let $\mathcal{E}$ denote the event that $ \sqrt{n}<|V_i|<(1+2\eps_2)\sqrt{n}$, for all $i\in[1,m]$.
Similarly as in the proof of \cref{lem:OPT-better}, the Chernoff bound implies that Event $\mathcal{E}$ occurs a.a.s.
We condition on the occurrence of $\mathcal{E}$.

Next, we show that $T$ is an $\alpha$-approximate traveling salesman tour, a.a.s.
For each $i\in[1,m-1]$, the distance between any point in $T_i$ and any point in $T_{i+1}$ is at most $\sqrt{5} \cdot D$, since $A_i$ and $A_{i+1}$ are adjacent squares, both of side length $D$.
The distance between the depot and any point in $[0,1]^2$ is $O(1)$ since the depot is fixed.
Therefore,
\begin{equation*}
\cost(T)\leq \left(\sum_{i=1}^m \cost(T_i)\right)+(m-1)\sqrt{5}\cdot D+O(1).
\end{equation*}
Using the same argument from \cref{sec:rect-I-II}, we obtain
\begin{equation*}
\sum_{i=1}^m\cost(T_i)<\left(1+\frac{\eps_0}{3}\right)\beta\sqrt{n},\quad\text{a.a.s}.
\end{equation*}
Noting that $(m-1)\sqrt{5}\cdot D=o(\sqrt{n})$, we have
\[\cost(T)<\left(1+\frac{\eps_0}{2}\right)\beta\sqrt{n}, \quad\text{a.a.s.}\]
On the other hand, by \cref{lem:TSP}, the value of an optimal traveling salesman tour on $V\cup\{O\}$ is at least $\left(1-\frac{\eps_0}{4}\right)\beta\sqrt{n}$, a.a.s.
Thus the approximation ratio of the solution $T$ is at most $1+\eps_0$ using the fact $\eps_0=\alpha-1\leq 1$.
Hence  $T$ is an $\alpha$-approximate traveling salesman tour on $V\cup\{O\}$, a.a.s.

It remains to lower bound $\ITP(T)$.
We say that a pair of points $(u,v)$ is a \emph{splitting pair} if $u$ and $v$ are consecutive points in $T$ and that $u$ is the ending point of one tour and $v$ is the starting point of another tour according to the construction in the ITP algorithm.
Let $w(u,v)$ denote the \emph{weight} of a splitting pair $(u,v)$  defined by $w(u,v)=\ell(u)+\ell(v)-\delta(u,v)$.
Note that $w(u,v)\geq 0$ by the triangle inequality.
From the definition of the $\ITP$ algorithm, the cost of the output solution satisfies
\begin{equation}
\label{eqn:cost-ITP}
\ITP(T)=\cost(T)+\SP,
\end{equation}
where $\SP$ denotes the total weight of the splitting pairs.

Observe that $\cost(T)$ is at least the overall cost of each $T_i$.
Again, using the same argument from \cref{sec:rect-I-II}, we obtain
\begin{equation}
\label{eqn:cost-TSP}
\cost(T)\geq \sum_{i=1}^m\cost(T_i)>(1-\eps_1)\beta\sqrt{n},\quad\text{a.a.s}.
\end{equation}

Next, we analyze the total weight $\SP$ of the splitting pairs.
For each $i\in[1,m]$, Event $\mathcal{E}$ implies that $|V_i|>\sqrt{n}$, so there is at least one splitting pair in $V_i$.
Let $(u_i,v_i)$ denote a splitting pair in $V_i$, breaking ties arbitrarily.
We have
$\SP\geq\sum_{i=1}^m w(u_i,v_i).$

Fix some $i\in[1,m]$.
In order to bound $w(u_i,v_i)$, we consider any point $x\in V_i$.
Since the three points $u_i$, $v_i$, and $x$ belong to the same square $A_i$ of side length $D$, their pairwise distances are at most $\sqrt{2}\cdot D$.
Thus both $\ell(u_i)$ and $\ell(v_i)$ are at least $\ell(x)-\sqrt{2}\cdot D$ by the triangle inequality.
Hence  \[w(u_i,v_i)=\ell(u_i)+\ell(v_i)-\delta(u_i,v_i)\geq 2\ell(x)-3\sqrt{2}\cdot D.\]
Averaging over all points $x\in V_i$ and using $|V_i|<(1+2\eps_2)\sqrt{n}$ (Event $\mathcal{E}$), we have
\begin{equation}
\label{eqn:w-ui-vi}
w(u_i,v_i)> \frac{2}{(1+2\eps_2)\sqrt{n}}\left(\sum_{x\in V_i} \ell(x)\right) - 3\sqrt{2}\cdot D.
\end{equation}
Summing \cref{eqn:w-ui-vi} over all $i\in[1,m]$ and recalling that $\rad=\frac{2}{\sqrt{n}}\sum_{x\in V}\ell(x)$, we have
\begin{equation}
\label{eqn:cost-SP}
\SP\geq \sum_{i=1}^m w(u_i,v_i)>\frac{1}{1+2\eps_2}\cdot \rad-3\sqrt{2}\cdot D\cdot m=\frac{1-o(1)}{1+2\eps_2}\cdot \rad>(1-\eps_1)\cdot\rad,
\end{equation}
where the equality follows from the facts that $\rad=\Theta(\sqrt{n})$ and $D\cdot m=o(\sqrt{n})$, and the last inequality follows from $\eps_2\leq \frac{\eps_1}{4}$.

The lower bound on $\ITP(T)$ in the claim follows from \cref{eqn:cost-ITP,eqn:cost-TSP,eqn:cost-SP}.

\subsection{Proof of \cref{thm:not-ptas}}
\label{sec:proof-not-ptas}
Let $\eps_1>0$ be a constant to be set later.
From \cref{lem:OPT-better}, there exists an absolute constant $c_1\in(0,\beta)$, such that $\OPT<(1+\eps_1)(\rad+\beta \sqrt{n})-c_1\sqrt{n}$, a.a.s.
From \cref{lem:ITP-tight}, for any $\alpha>1$, there exists an $\alpha$-approximate traveling salesman tour $T$ on $V\cup\{O\}$, such that $\ITP(T)>(1-\eps_1)(\rad+\beta \sqrt{n})$, a.a.s.
Hence
\[
\frac{\ITP(T)}{\OPT}>\frac{(1-\eps_1)(\rad+\beta\sqrt{n})}{(1+\eps_1)(\rad+\beta\sqrt{n})-c_1\sqrt{n}},\quad\text{a.a.s.}
\]
To analyze $\rad$, let $L$ be the expectation of $\ell(x)$ for $x\in[0,1]^2$ with uniform distribution.
Since the depot $O$ is at a constant distance from $[0,1]^2$, $L$ is a constant.
By the law of large numbers,
\[\bigg|\frac{1}{n}\sum_{x\in V}\ell(x)-L\bigg|<\eps_1,\quad\text{a.a.s.}\]
Recall that $\rad=\frac{2}{\sqrt{n}}\sum_{x\in V}\ell(x)$.
Hence $(L-\eps_1)\cdot 2\sqrt{n}<\rad<(L+\eps_1)\cdot 2\sqrt{n}$, a.a.s.
Denoting the function $f$ as
\[ f(\eps_1)=\frac{(1-\eps_1)(2L-2\eps_1+\beta)}{(1+\eps_1)(2L+2\eps_1+\beta)-c_1},\]
we have
\[
\frac{\ITP(T)}{\OPT}>f(\eps_1),\quad\text{a.a.s.}
\]
Since $L$, $\beta$ and $c_1$ are positive constants and $c_1<\beta$ (\cref{lem:OPT-better}), we have
\[\lim_{\eps_1\to 0}f(\eps_1)=1+\frac{c_1}{2L+\beta-c_1}.\]
Let $c_0=\frac{1}{2}\cdot\frac{c_1}{2L+\beta-c_1}$, which is a positive constant.
Choosing $\eps_1$ small enough such that $f(\eps_1)>1+c_0$, we have
\[\frac{\ITP(T)}{\OPT}>f(\eps_1)>1+c_0,\quad\text{a.a.s.}\]

We complete the proof of \cref{thm:not-ptas}.

\section{Upper Bound on the Approximation Ratio}
\label{sec:better-ratio}
In this section, we prove \cref{thm:better-ratio} by providing an upper bound on the approximation ratio $\ITP(T)/\OPT$ of the ITP algorithm, where $T$ is a traveling salesman tour.

Let $\lambda$ and $\eps$ be positive constants such that $\lambda+\eps<1$.
We analyze the performance of the ITP algorithm with respect to $\lambda$ and $\eps$.
The values of $\lambda$ and $\eps$ will be set in the end of the proof.

\subsection{Structural analysis}

\begin{lemma}
\label{lem:single-tour}
Let $T=(O,y_1,y_2,\dots, y_m,O)$ be any tour starting and ending at $O$.
Let $L=\frac{1}{m}\left(\sum_{j=1}^m \ell(y_j)\right)$.
Let $\Delta=\max_{1\leq j\leq m} \{\ell(y_j)\}-L$.
Then there exists a set $W\subseteq \{y_1,\dots,y_m\}$ which is of cardinality greater than $(\lambda+\eps) \cdot m-1$ such that
\begin{equation*}
\cost(T)\geq 2\left(L-\frac{\lambda+\eps}{1-\lambda-\eps}\cdot\Delta\right)+\sum_{x\in W} \delta(x,W\setminus\{x\}).
\end{equation*}
\end{lemma}

\begin{proof}
Let $W$ denote the set of points $x$ such that $\ell(x)\geq L-\frac{\lambda+\eps}{1-\lambda-\eps}\cdot\Delta$, other than the last one in the order of traversal by $T$ starting from $O$. Tour $T$ must first travel through a path to a first point of $W$, paying at least $L-\frac{\lambda+\eps}{1-\lambda-\eps}\cdot\Delta$, then proceed from each point $x$ of $W$ through a path to another point of $W$, paying at least $\delta(x,W\setminus\{x\})$, and finally, go to one more point such that $\ell(x)\geq L-\frac{\lambda+\eps}{1-\lambda-\eps}\cdot\Delta$, and travel from there through a path back to the depot, paying at least $L-\frac{\lambda+\eps}{1-\lambda-\eps}\cdot\Delta$.
Hence the cost of $T$ is at least as stated in Lemma~\ref{lem:single-tour}.

Next, we bound the size of $W$.
When $\Delta=0$, we have $\ell(y_j)=L$ for all $j\in[1,m]$. Hence $|W|=m-1$, which is greater than $(\lambda+\eps)\cdot m -1$, since $\lambda+\eps<1$.
The claim follows.

It remains to analyze the case when $\Delta>0$.
Every point of $T$ is at distance at most $L+\Delta$ from the depot. Letting $m'$ denote the number of points whose distance from the depot is at least $L-\frac{\lambda+\eps}{1-\lambda-\eps}\cdot\Delta$, we have
\[mL= \sum_{j=1}^m \ell(y_j) < m'(L+\Delta)+(m-m')\left(L-\frac{\lambda+
\eps}{1-\lambda-\eps}\cdot\Delta\right).\]
Since  $1-\lambda-\eps>0$, this implies $m'-(\lambda+\eps)\cdot m> 0$, hence
$|W|=m'-1 > (\lambda+\eps)\cdot m-1$.
\end{proof}

The following result is a strengthening of the lower bound $\OPT\geq \rad$ from \cref{lem:OPT-lower-bound}, and will lead to our improved analysis of the ITP algorithm.
\begin{theorem}
\label{thm:new-lower-bound}
Let $\lambda$ and $\eps$ be positive constants such that $\lambda+\eps<1$.
Let $V$ be a set of $n$ points in any distance metric.
Let $k=\omega(1)$.
There exists a set $U\subseteq V$ which is of cardinality greater than  $\left(\lambda+\frac{\eps}{2}\right)\cdot n$ for $n$ large enough, and such that
\[\OPT\geq \rad+(1-\lambda-\eps) \left(\sum_{x\in U} \delta(x,U\setminus\{x\})\right).\]
\end{theorem}

\begin{proof}
Let $T_1, \dots, T_q$ be the tours in an optimal solution to the CVRP.
Let $m_i$ be the number of points in $V$ that are visited by the tour $T_i$.
Up to combining tours that visit few points, we may assume that $m_i> \frac{k}{2}$ for all but at most one tour, so $q\leq \frac{2n}{k}+1=o(n)$.

For each tour $T_i$, define the corresponding $L_i$, $\Delta_i$, and $W_i$ with respect to the tour $T_i$ using the notations of \cref{lem:single-tour}.
By summation, letting $U=\bigcup_i W_i$, Lemma~\ref{lem:single-tour} then implies (using $q=o(n)$, $n$ large enough, and $\delta(x,W_i\setminus\{x\})\geq \delta(x,U\setminus\{x\}$))
\[|U|=\sum_{i\leq q} |W_i|>\bigg((\lambda+\eps)\sum_i m_i\bigg) - q=(\lambda+\eps)\cdot n - q>\left(\lambda+\frac{\eps}{2}\right)\cdot n\]
and
\begin{equation}
\label{eqn:1}
\sum_i \cost(T_i)\geq \left(\sum_i 2\left(L_i-\frac{\lambda+\eps}{1-\lambda-\eps}\cdot\Delta_i\right)\right)+\left(\sum_{x\in U} \delta(x,U\setminus\{x\})\right).
\end{equation}

On the other hand, we trivially have
\begin{equation}
\label{eqn:2}
\sum_i \cost(T_i)\geq \sum_i 2(L_i+\Delta_i).
\end{equation}

A linear combination of \cref{eqn:1} with coefficient $(1-\lambda-\eps)$ and of
\cref{eqn:2} with coefficient $(\lambda+\eps)$ leads to:
\[\OPT= \sum_i \cost(T_i)\geq \left(\sum_i 2L_i\right) + (1-\lambda-\eps)\left(\sum_{x\in U} \delta(x,U\setminus\{x\})\right).\]

Observe that \[\sum_i 2L_i=\sum_i \sum_{x\in T_i} \frac{2\ell(x)}{m_i}\geq \sum_i \sum_{x\in T_i} \frac{2\ell(x)}{k}=\sum_{x\in V} \frac{2\ell(x)}{k}=\rad.\] The Lemma follows.
\end{proof}

\subsection{Probabilistic Analysis}
The following result suggests that the closest point distance follows  the law of large numbers. It is a corollary of Theorem~2.4 in~\cite{penrose2003weak}.\footnote{
To apply Theorem~2.4 in~\cite{penrose2003weak}, we consider a \emph{directed} graph $G$ with vertex set $V$,  such that from every vertex $x\in V$, there is a unique outgoing edge, let it be $(x,y)$, where $y$ is the closest point to $x$ among the points in $V\setminus\{x\}$, breaking ties arbitrarily.
Theorem~2.4 in~\cite{penrose2003weak} is interpreted with reference to Remark (h) in~\cite{penrose2003weak}.}
\begin{lemma}[\cite{penrose2003weak}]
\label{lem:large-numbers}
Let $\mathcal{P}$ be a homogeneous Poisson point process of intensity 1 on $\mathbb{R}^2$ and $\delta_\mathcal{P}$ denote the distance from the origin of $\mathbb{R}^2$ to a closest point in $\mathcal{P}$ by the Euclidean norm. Let $V$ be a set of $n$ i.i.d.\ uniform random points. Then, given any bounded function $\phi: [0,\infty]\to [0,\infty)$, as $n\to\infty$ we have:
\[
\frac{1}{n} \sum_{x\in V} \phi\left(\sqrt{n}\cdot\delta(x,V\setminus\{x\})\right)
\to \E\big[\phi(\delta_\mathcal{P})\big].
\]
\end{lemma}

\cref{lem:large-numbers} provides the rigorous setting enabling us to derive a new lower bound on the sum of the closest point distances over a subset of a set of i.i.d.\ uniform random points, which we now state.
\begin{lemma}
\label{lem:random}
Let $V$ be a set of $n$ i.i.d.\ uniform random points.
Let $U$ be any subset of $V$ such that $|U|> \left(\lambda+\frac{\eps}{2}\right)\cdot n$.
Then, asymptotically almost surely,
\[\sum_{x\in U}\delta(x,V\setminus\{x\})> (\xi_\lambda-\epsilon)\cdot \sqrt{n},\]
where $\xi_\lambda$ is a constant defined by
\[\xi_\lambda:=\frac{1}{2}\erf\left(\sqrt{\ln \frac{1}{1-\lambda}}\right)-(1-\lambda)\cdot\sqrt{\frac{1}{\pi}\cdot\ln \frac{1}{1-\lambda}}\]
in which $\erf(\cdot)$ is the Gauss error function $\erf(z)=\frac{2}{\sqrt{\pi}}\int_{0}^{z}e^{-t^2} dt$.
\end{lemma}

\begin{proof}
Recall the definition of $\delta_\mathcal{P}$ from \cref{lem:large-numbers}.
By definition of the Poisson point process, the probability  $g(r)$ of the event $\delta_\mathcal{P}\leq r$ equals $1-e^{-\pi r^2}$.

Let $Z\subseteq V$ be the set of points $x\in V$ such that $\delta(x, V\setminus\{x\})\leq \frac{r_0}{\sqrt{n}}$, with $r_0=\sqrt{\frac{1}{\pi}\cdot\ln \frac{1}{1-\lambda}}$.
We apply \cref{lem:large-numbers} with $\phi$ equals $\phi_1$, the indicator function of whether $r\leq r_0$, to obtain that, as $n\to \infty$,
\[\frac{|Z|}{n}\to \E\big[\phi_1(\delta_\mathcal{P})\big]=g(r_0)=\lambda.\]
Thus $|Z|\leq \left(\lambda+\frac{\eps}{2}\right)\cdot n< |U|$, a.a.s.
Since $Z$ consists of the points $x\in V$ with the smallest values of $\delta(x, V\setminus\{x\})$, we have
\begin{equation}
\label{eqn:hat-X}
\sum_{x\in U} \delta(x, V\setminus\{x\})\geq
\sum_{x\in Z} \delta(x, V\setminus\{x\}),\quad\text{a.a.s.}
\end{equation}
To analyze $\displaystyle \sum_{x\in Z} \delta(x, V\setminus\{x\})$, we define a bounded function $\phi_2$ as follows.
\[
\phi_2(r)=
\begin{cases}
r, & r\leq r_0\\
0, & \text{otherwise}.
\end{cases}
\]
Applying \cref{lem:large-numbers} with $\phi=\phi_2$, we have, as $n\to\infty$,
\[\frac{1}{n}\displaystyle \sum_{x\in Z} \sqrt{n}\cdot\delta(x, V\setminus\{x\})\to \E\big[\phi_2(\delta_\mathcal{P})\big],\]
thus
\begin{equation}
\label{eqn:Z}
\sum_{x\in Z} \delta(x, V\setminus\{x\})> (\E\big[\phi_2(\delta_\mathcal{P})\big]-\eps)\cdot \sqrt{n},\quad\text{a.a.s.}
\end{equation}
Observe that $\E\big[\phi_2(\delta_\mathcal{P})\big]=\displaystyle\int_0^{\infty} \phi_2(r)\cdot g'(r)dr=\displaystyle\int_0^{r_0} r\cdot g'(r)dr$, where $g'(r)= 2\pi r\cdot e^{-\pi r^2}$.
Integrating by parts, recalling the definition of the Gauss error function, and plugging in the value of $r_0$, we have
\[\E\big[\phi_2(\delta_\mathcal{P})\big]
=\left(\int_0^{r_0} e^{-\pi r^2} dr\right) -\bigg[
r \cdot e^{-\pi r^2}\bigg]_0^{r_0}
=\left[
\frac{\erf(\sqrt{\pi}\cdot r)}{2}-r \cdot e^{-\pi r^2}\right]_0^{r_0}=\xi_\lambda.
\]
The claim follows.
\end{proof}

\subsection{Proof of \cref{thm:better-ratio}}
Let $T$ denote an $\alpha$-approximate traveling salesman tour on $V\cup\{O\}$, where $\alpha\geq 1$ is a constant.
When $k=O(1)$, for any $\eps>0$, $\ITP(T)<(1+\eps)\OPT$ a.a.s.\ according to \cite{haimovich1985bounds}, which implies the claim.
In the following, we assume that $k=\omega(1)$.

According to \cref{lem:ITP-upper-bound}, we have \[\ITP(T)< \cost(T)+\rad.\]

First, we bound $\cost(T)$.
Letting $T^*$ denote an optimal traveling salesman tour on $V\cup\{O\}$, we have $\cost(T)\leq \alpha\cdot \cost(T^*)$.
By \cref{lem:TSP}, the value of an optimal traveling salesman tour on $V$ is less than $\left(\beta+\frac{\eps}{2}\right)\cdot\sqrt{n}$, a.a.s.
Since the distance between the depot and any point in $[0,1]^2$ is $O(1)$, we have $\cost(T^*)< \left(\beta+\frac{\eps}{2}\right)\cdot\sqrt{n}+O(1)$, which is less than $(\beta+\eps)\cdot\sqrt{n}$ when $n$ is large enough.
Thus $\cost(T)<\alpha (\beta+\eps)\cdot\sqrt{n},$  a.a.s.

Next, we analyze $\rad$.
By \cref{thm:new-lower-bound}, for some $U\subseteq V$ of size greater than $\left(\lambda + \frac{\epsilon}{2}\right)n$ we have
\[\rad\leq \OPT-(1-\lambda-\eps) \left(\sum_{x\in U} \delta(x,U\setminus\{x\})\right).\]
By \cref{lem:random} and the fact that $\delta(x,U\setminus\{x\})\geq \delta(x,V\setminus\{x\})$,  we have a.a.s.
\[\sum_{x\in U} \delta(x,U\setminus\{x\}) > (\xi_\lambda-\eps)\cdot \sqrt{n}.\]
Noting that $1-\lambda-\eps>0$, we have
\[\rad\leq \OPT-(1-\lambda-\eps) \cdot (\xi_\lambda-\eps)\cdot \sqrt{n}.\]

\begin{figure}[t]
\includegraphics[scale=0.3]{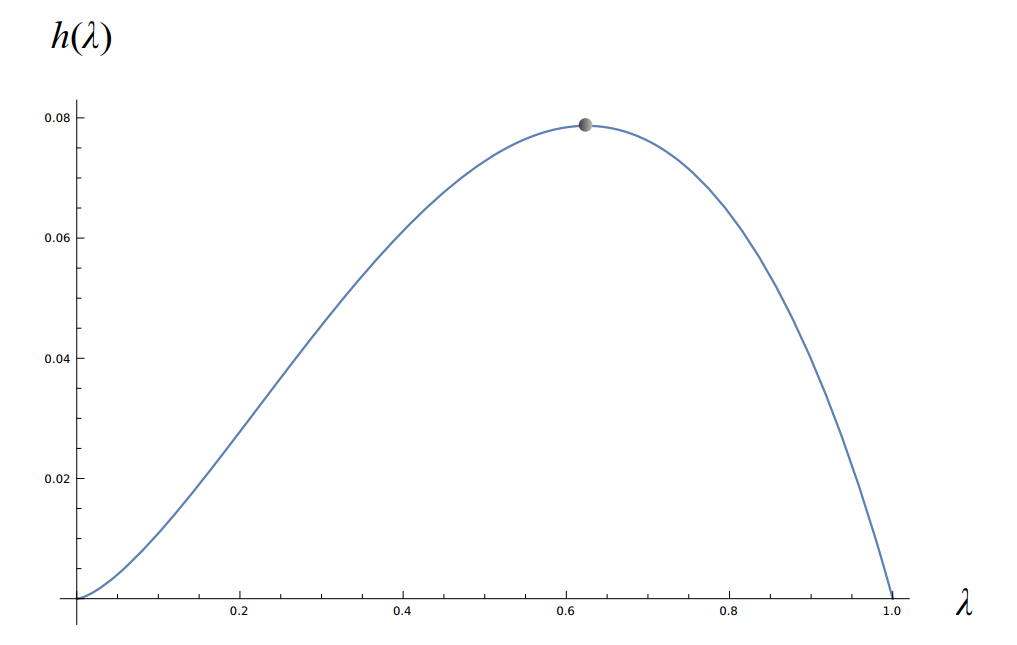}
\caption{Plot of the function $h(\lambda)=(1-\lambda )\cdot \xi_\lambda$ for $\lambda\in[0,1)$. The maximum value of $h(\lambda)$ is greater than $0.078674$, which is achieved when $\lambda$ is roughly 0.62468.}
\label{fig:plot}
\end{figure}

Combining the above bounds gives a.a.s.
\begin{equation}
\label{eqn:SOL}
\ITP(T)< \OPT+ \big(\alpha(\beta+\eps)-(1-\lambda-\eps)\cdot (\xi_\lambda-\eps)\big)\cdot \sqrt{n}.
\end{equation}
Note that the coefficient of $\sqrt{n}$ in \cref{eqn:SOL} must be positive, because $\ITP(T)\geq \OPT$ (\cref{lem:ITP-upper-bound}).
Using \cref{lem:TSP,lem:OPT-lower-bound}, and assuming $\eps<\beta$, we have a.a.s.
$\sqrt{n}< \frac{\cost(T^*)}{\beta-\eps} \leq \frac{\OPT}{\beta-\eps},$
and substituting into  \cref{eqn:SOL} gives a.a.s.
\[\ITP(T)< \left(1 + \frac{\alpha(\beta+\eps)-(1-\lambda-\eps)\cdot (\xi_\lambda-\eps)}{\beta-\eps}\right)\cdot\OPT.\]
Since $\beta$ is a positive constant (\cref{lem:TSP}), choosing $\lambda$ to maximize $(1-\lambda)\cdot \xi_\lambda$ and $\eps$ small enough yields
\[\frac{\ITP(T)}{\OPT}<1+\alpha-\frac{\max_\lambda \big\{(1-\lambda)\cdot \xi_\lambda\big\}}{\beta}+0.00001.\]
A numerical calculation (\cref{fig:plot}) gives $\max_{\lambda} \big\{(1-\lambda)\cdot \xi_\lambda\big\}> 0.078674$, and \cref{lem:TSP} tells us that $\beta<\beta_1=0.92117$. Substituting those values concludes the proof.

\subsection{Generalization to Multiple Depots}
\label{sec:multi-depot}

In this section, we generalize \cref{thm:better-ratio} to the setting of multiple depots.

Let $S$ be a set of $s\in \mathbb{N}$ depots, where the number $s$ and the elements in $S$ are fixed in advance.
We require that each route starts and ends at the same depot in $S$, and visits at most $k$ customers.
As before, the objective is to find a collection of routes of minimum total cost, so that each customer is visited by some route.
We assume that there is no restriction on the number of vehicles available at each depot.

We consider the generalized ITP algorithm in the setting of multiple depots, due to Bompadre, Dror, and Orlin~\cite{bompadre2007probabilistic}.
The algorithm first assigns each customer to the closest depot and then solves $s$ independent subproblems using the ITP algorithm for a single depot.
Let $T$ denote the union of the $s$ traveling salesman tours in the ITP algorithm on those subproblems.
Assume that each traveling salesman tour is an $\alpha$-approximation to the corresponding subproblem, for some constant $\alpha>1$.
Let $\ITP(T)$ denote the total cost of the solutions to the $s$ subproblems obtained by the ITP algorithm.

We generalize the definition of the function $\ell$, previously defined in the setting of a single depot. For any $x\in V$, let $\ell(x)=\min_{O\in S} \{\delta(O,x)\}$.
The radial cost is again defined by $\rad=\frac{2}{k}\cdot \sum_{x\in V} \ell(x)$.

When $k=O(1)$, for any $\eps>0$, $\ITP(T)<(1+\eps)\OPT$ a.a.s.\ according to \cite{bompadre2007probabilistic}, which implies the claim.
In the following, we assume that $k=\omega(1)$.

Our main observation here is that \cref{thm:new-lower-bound} holds in the setting of multiple depots.
Indeed, the proof of \cref{thm:new-lower-bound} proceeds in the same way, except that the number of tours is now bounded by at most $\frac{2n}{k}+s$, which is again $o(n)$ since $s$ is fixed and $k=\omega(1)$.

To extend \cref{thm:better-ratio} to the setting of multiple depots, we also need several results by Bompadre, Dror, and Orlin~\cite{bompadre2007probabilistic}, which we summarize below.
According to Lemma~8 in~\cite{bompadre2007probabilistic},
\[\OPT\leq \ITP(T)\leq \rad+\cost(T).\]
Let $T^*$ be an optimal traveling salesman tour on $V\cup S$.
By Lemma~15 in~\cite{bompadre2007probabilistic}, \[\cost(T)\leq \alpha\cdot\cost(T^*),\quad\text{a.a.s.}\]
By Lemma~10 in~\cite{bompadre2007probabilistic} and observing that the cost of an optimal traveling salesman tour on $S$ is $O(1)$ (since the depots are fixed), we have
\[\cost(T^*)<\OPT+O(1).\]
The claim of \cref{thm:better-ratio} in the setting of multiple depots follows, by substituting all the elements above into the initial version of the proof in the setting of a single depot.


\bibliography{references}

\end{document}